\documentclass{ws-p8-50x6-00}

\usepackage{psfig}

\begin{document}
\psfigurepath{.}

\title{Magnetic Properties of Heavy Fermion Superconductors CeRhIn$_5$ and Ce$_2$RhIn$_8$}

\author{Wei Bao$^1$, G. Aeppli$^2$, A.D. Christianson$^1$, Z. Fisk$^{3,1}$,\\ 
M.F. Hundley$^1$, A.H. Larcerda$^1$, J.W. Lynn$^4$,\\ P.G. Pagliuso$^1$, 
J.L. Sarrao$^1$, J.D. Thompson$^1$}

\address{$^1$Los Alamos National Laboratory, Los Alamos, NM 87545, USA\\  
$^2$NEC Research, 4 Independence Way, Princeton, NJ 08540, USA\\
$^3$Florida State University, Tallahassee, FL 32306, USA\\
$^4$National Institute of Standards and Technology, Gaithersburg, MD 20899, USA}


\maketitle

\abstracts{Some recent neutron scattering works on
CeRhIn$_5$ and Ce$_2$RhIn$_8$, together with
related resistivity and specific heat measurements, are summarized.
In spite of its layered crystal structure, CeRhIn$_5$ is shown
to be 3 dimensional both magnetically and in transport.
We also find that the Fisher-Langer behavior is closely followed
in CeRhIn$_5$. This may circumvent the Kondo lattice model
and support applying established Fermi-liquid superconductivity theory to
heavy fermion superconductors.
}

\section{Introduction}

Three materials, Ce$M$In$_5$ ($M$=Rh, Ir, Co), of the same 
HoCoGa$_5$ crystal structure
recently have been added to the list of Ce-based heavy fermion superconductors\cite{hegger,JoeIr,JoeCo}.
Previously, the list contained only one ambient pressure superconductor,
CeCu$_2$Si$_2$ (T$_C=0.7$~K)\cite{steg}. The others, CeCu$_2$Ge$_2$
(T$_C=0.64$~K at 10 GPa)\cite{djkb}, 
CePd$_2$Si$_2$ (T$_C=0.5$~K at 2.5 GPa)\cite{roma},
CeRh$_2$Si$_2$ (T$_C=0.35$~K at 0.9 GPa)\cite{gros} and CeIn$_3$ 
(T$_C=0.2$~K at 2.5 GPa)\cite{walk}, 
superconduct only under high pressures. Of the three new materials, the Ir and
Co compounds are ambient pressure superconductors with T$_C= 0.4$~K and
2.3~K respectively, while the Rh compound superconducts at 2.1~K
above 1.6 GPa\cite{Sarrao}.

Ce$M$In$_5$ is structurally related to the previously known superconductor
CeIn$_3$ by alternately stacking CeIn$_3$ and $M$In$_2$ layers.
Particularly, it is interesting to compare CeIn$_3$ and CeRhIn$_5$, 
both of which are antiferromagnets at ambient pressure. The optimal 
T$_C$ of the layered compound is 10 times that for cubic
CeIn$_3$. In terms of the strength of magnetic interactions as 
represented by the N\'{e}el temperature, T$_C$/T$_N$ of CeRhIn$_5$ 
is 28 times that for CeIn$_3$. In term of the Fermi energy 
via the Sommerfeld
constant, the enhancement of T$_C$/E$_f \sim $T$_C \gamma$ from CeIn$_3$ 
to CeRhIn$_5$ is an even more impressive value of 35. 
Monthoux and Lonzarich recently argued that 2-dimensional (2D) magnetic 
fluctuations are superior to 3D magnetic fluctuations in elevating 
T$_C$\cite{pmggl1}.
It is, therefore, natural to ask whether this mechanism is working
for Ce$M$In$_5$. de Haas-van Alphen (dHvA) measurements on CeRhIn$_5$, 
CeIrIn$_5$ and CeCoIn$_5$ have provided
evidence for the existence of 2D Fermi sheets in
addition to 3D ones\cite{haga,hall_Rh,hall_Co}. Aspects of magnetic measurements, such
as a small $\beta$ critical exponent, have been used to suggest magnetic
2-dimensionality. However, we are going to show with
neutron scattering and bulk measurements that at least for
CeRhIn$_5$, both magnetic and transport properties are 3D\cite{bao01b}.
Furthermore, the close relation among resistivity, specific heat and 
antiferromagnetic fluctuations in the neighborhood of T$_N$,
in the fashion illustrated by Fisher and Langer, indicates
the Kondo lattice of CeRhIn$_5$ has been renormalized at
low temperatures to weakly coupled subsystems of local magnetic moments 
and heavy fermions. If this holds true at the critical pressure for CeRhIn$_5$
and also true for CeIrIn$_5$ and CeCoIn$_5$, one can bypass 
theoretical difficulties possed by the Kondo lattice model, 
and directly treat
superconductivity in these new heavy fermion materials with a Fermi
liquid model using the experimentally measured magnetic fluctuation
spectra for the bosons.

\section{Experimental Results and Discussions}

\begin{figure}[bh]
\centerline{
\psfig{file=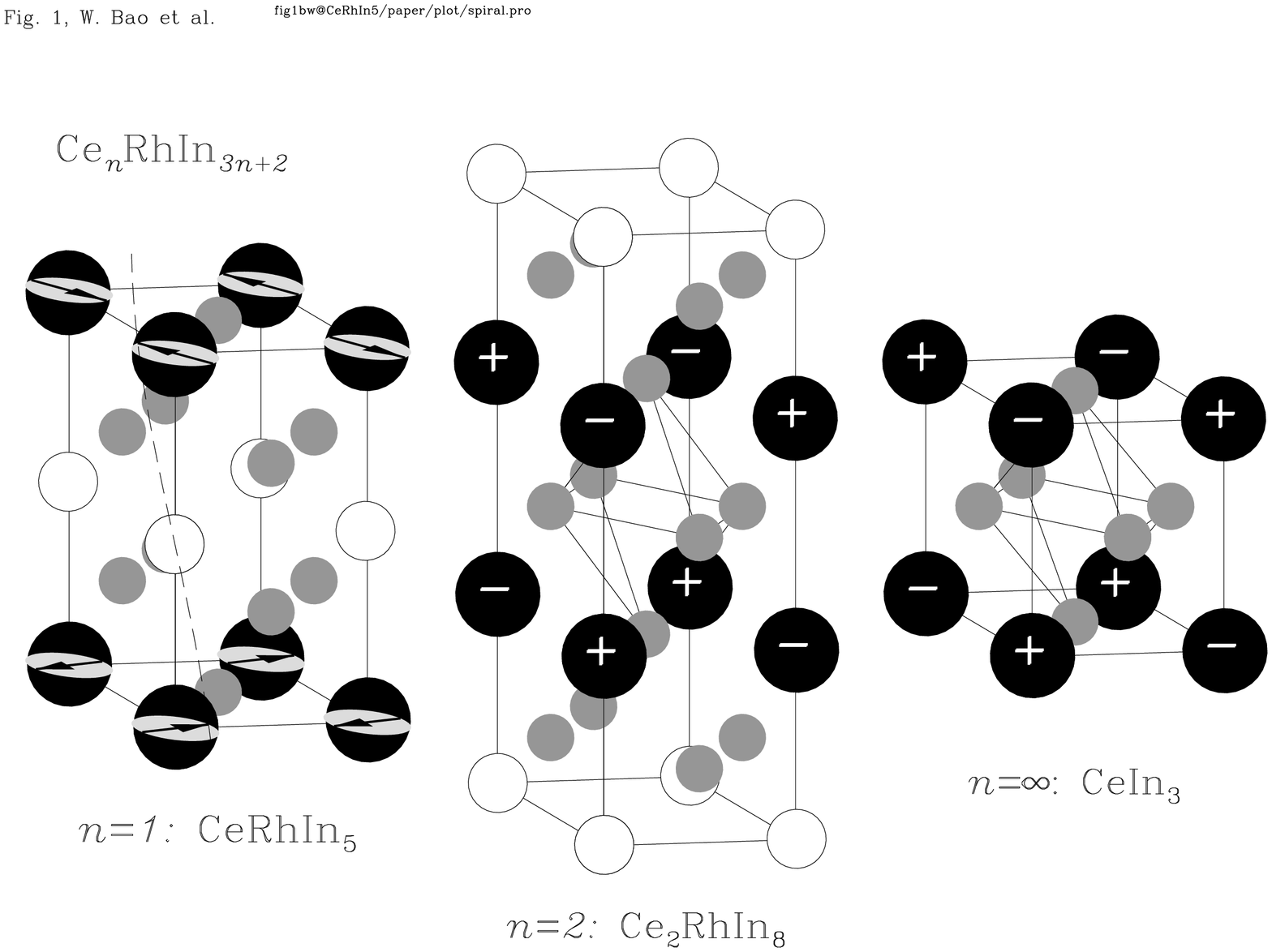,width=.6\columnwidth,angle=90,clip=}}
\vskip -.8cm
\caption{Magnetic structures of Ce$_n$RhIn$_{3n+2}$. The solid circle
denotes Ce with arrow or signs indicating moment orientation. The shaded
circle denotes In, the open circle Rh. Magnetic moment per Ce is
0.37$\mu_B$ for CeRhIn$_5$\protect\cite{bao00a}, 
0.55$\mu_B$ for Ce$_2$RhIn$_8$\protect\cite{bao01a}
and 0.48-65$\mu_B$ for CeIn$_3$\protect\cite{cein,ssc}.
From ref.\ [\protect\cite{bao01a}].}
\end{figure}
Fig.\ 1 shows magnetic structures of Ce$_n$RhIn$_{3n+2}$ (n=1, 2 
and $\infty$)
determined with neutron diffraction. Notice that the nearest
neighbor (n.n.) antiferromagnetic pairs of the three materials 
have identical local environments.
The pairs separated by the RhIn$_2$ layer are collinear for n=2
but are incommensurate for n=1. The incommensurate magnetic structure
of CeRhIn$_5$ is found to transform to a commensurate one by
a 2-3 T magnetic field applied in the layer at low T\cite{bao02b}. 
An additional
incommensurate antiferromagnetic component appears below 1.4~K
for Ce$_2$RhIn$_8$. They suggest competing magnetic interaction between 
the Ce pairs. 
The $T$-$H$ phase diagram also reveals the N\'{e}el point as a
multicritical point\cite{bao02b}. This explains the small $\beta$ critical
exponent.

Now let us address the dimensionality issue.
Fig.\ 2 shows antiferromagnetic fluctuations along the $c$-axis
for CeRhIn$_5$.
\begin{figure}[b]
\centerline{
\psfig{file=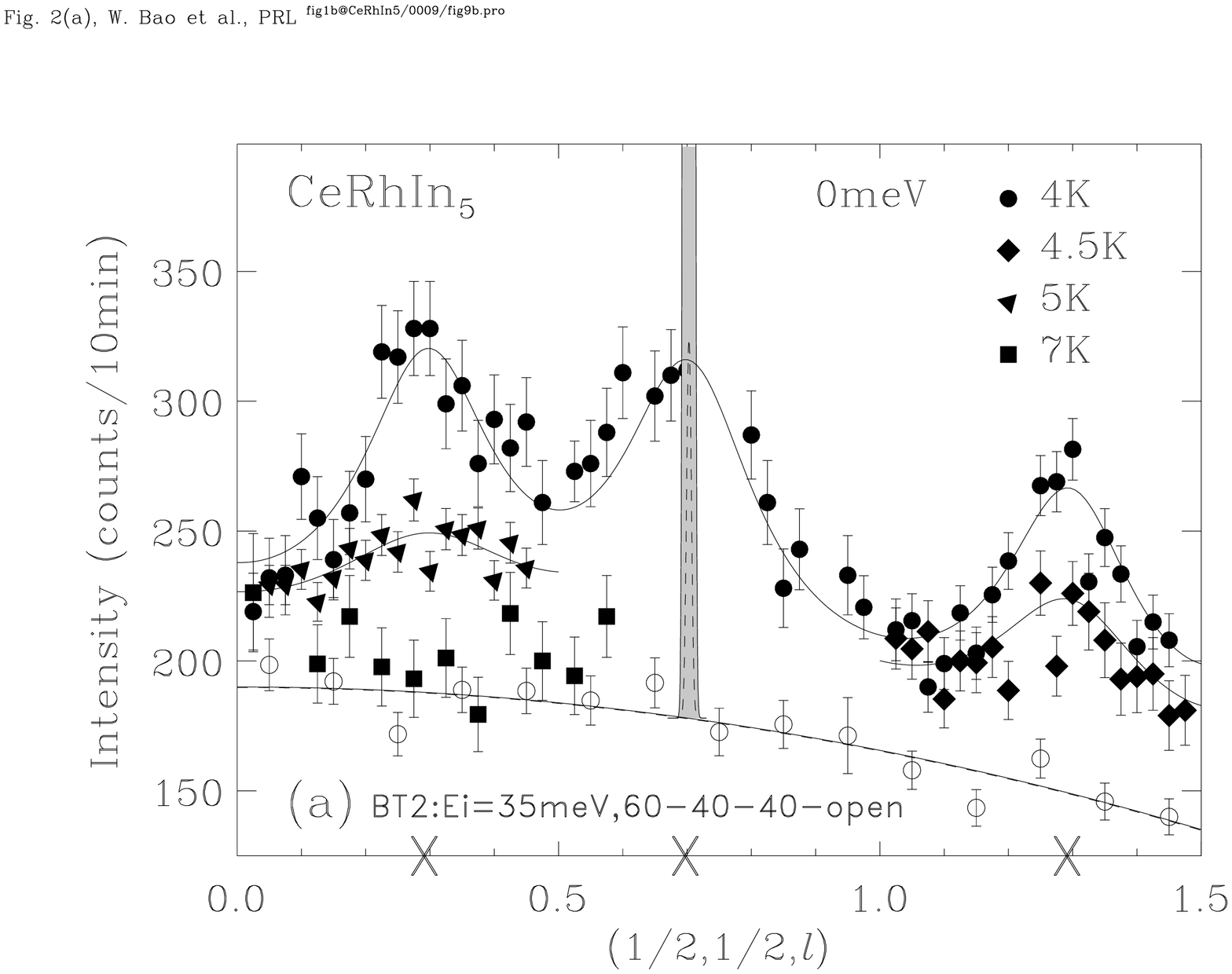,width=.6\columnwidth,angle=90,clip=}}
\vskip -.5cm
\caption{Instantaneous magnetic correlation function ${\bf S}({\bf q})$
measured along the $c$ axis at various temperatures. The open circles
indicate background. From ref.\ [\protect\cite{bao01b}].}
\end{figure}
There exists strong intensity modulation in step with
magnetic Bragg peaks, which are marked by the crosses. 
This directly contradicts the idea that CeRhIn$_5$
is a formally 2D magnetic system.
Magnetic correlation lengths along the $c$ axis and
in-plane have the same order of magnitude and evolve with temperature
in a similar fashion\cite{}. This also contradicts a 2D magnetic
model for CeRhIn$_5$. From the correlation lengths, it is deduced 
that the magnetic interaction
for the further separated n.n.\ Ce pairs along the $c$ axis is weaker than
that for the n.n. Ce pairs along the $a$ axis, consistent with a robust
antiferromagnetic bond for the n.n. Ce pairs in the plane. 

Fig.\ 3 shows resistivity measured with current in-plane and along 
the $c$ axis for CeRhIn$_5$.
\begin{figure}[bt]
\centerline{
\psfig{file=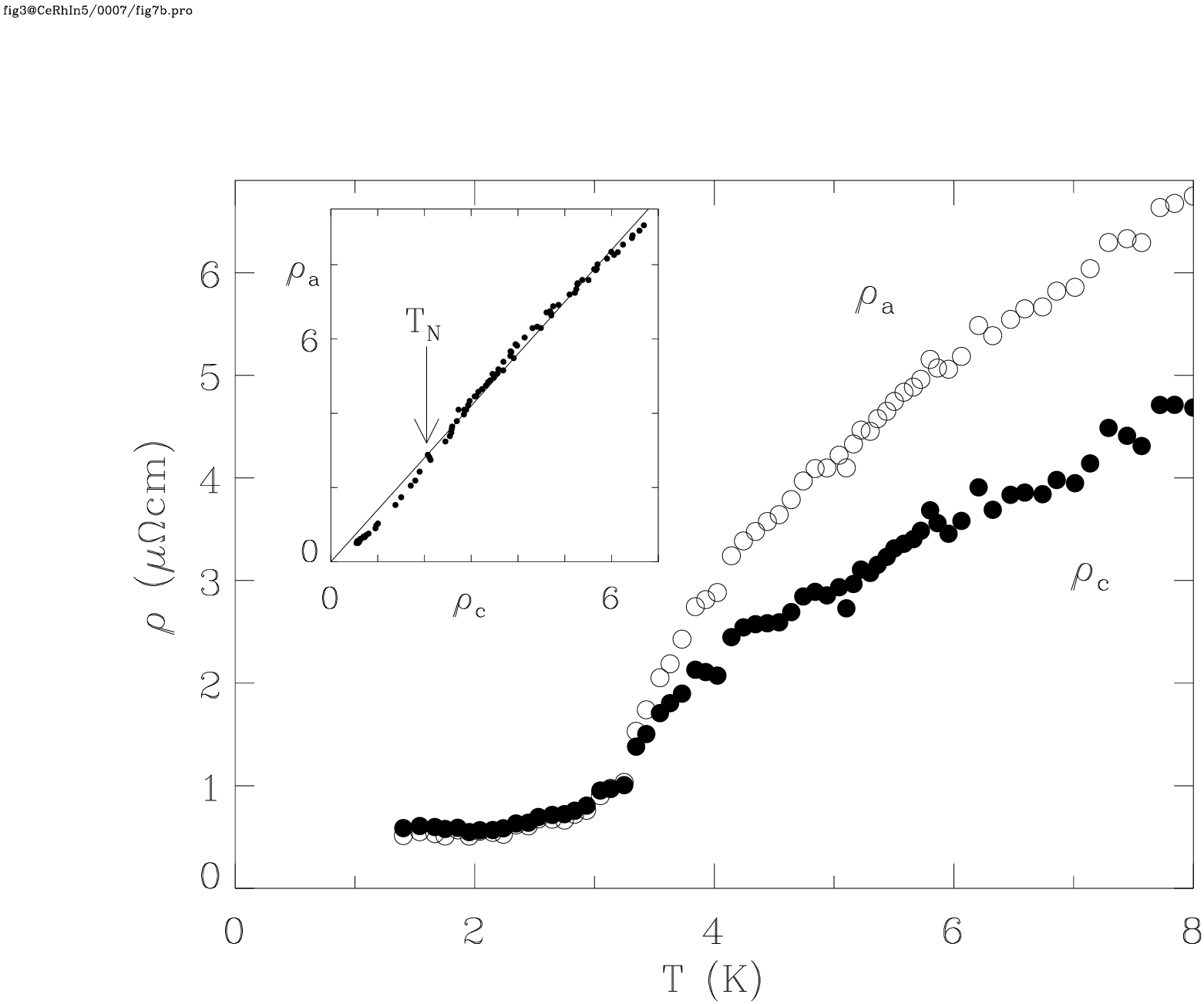,width=.5\columnwidth,angle=90,clip=}}
\vskip -.5cm
\caption{Resistivity measured with current along the $a$ axis and the
$c$ axis respectively. From Ref.\ [\protect\cite{bao01b}].}
\end{figure}
At lowest temperatures, resistivity out of plane, $\rho_c$,
is comparable to the in-plane $\rho_a$. At higher temperatures,
$\rho_c$ is even smaller than $\rho_a$. This
clearly rules out for CeRhIn$_5$ 2D electronic transport, which
requires $\rho_c \gg \rho_a$.
Although there is 2D band at the Fermi energy as revealed in 
the dHvA experiments, 3D bands clearly dominate in transport.

The single impurity Kondo model of N electrons has very complex behavior
at finite temperatures. It however renormalizes to a simple N-1 electron system
at T=0. It is also possible for the Kondo lattice
model, which describes a heavy fermion system, to renormalized to 
a simple system at low T. But a reliable theoretical
prediction for possible states in a real material is difficult.
Fig.\ 4 compares the intensity of antiferromagnetic fluctuations at 
magnetic Bragg point,
derivative of resistivity, and magnetic part of specific heat for CeRhIn$_5$.
\begin{figure}[bt]
\centerline{
\psfig{file=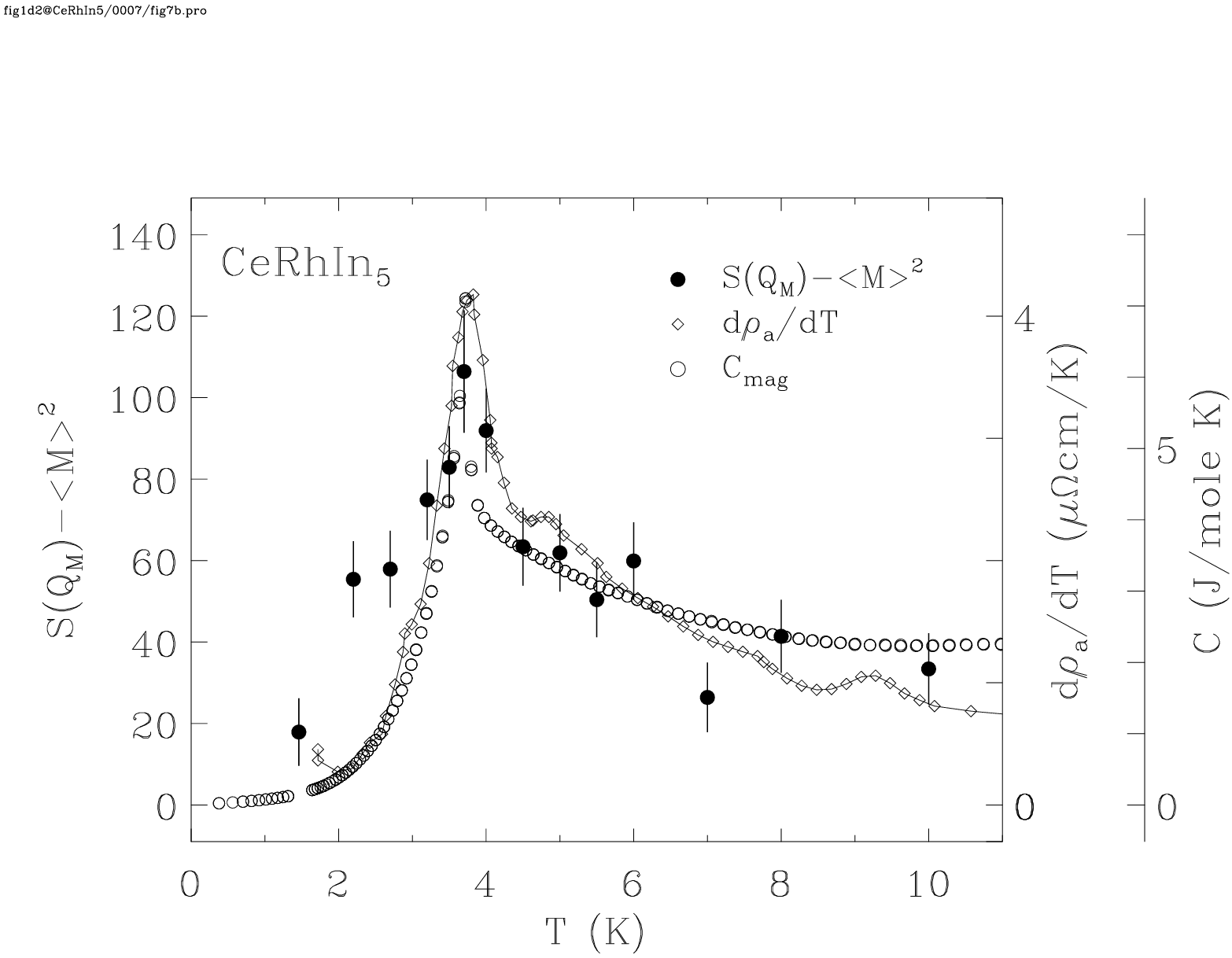,width=.6\columnwidth,angle=90,clip=}}
\vskip -.6cm
\caption{Intensity of magnetic fluctuations 
${\bf S}({\bf Q}_M)-\langle M\rangle ^2$, derivative of resistivity, and magnetic specific heat as a function of temperature.
From Ref.\ [\protect\cite{bao01b}].}
\end{figure}
These quantities are remarkably similar. 
Fisher and Langer\cite{fisherl} have considered
a model which can be regarded as the Kondo lattice model at the small
Kondo interaction limit and which can be solved with the Born approximation. 
They predict the behavior for the three quantities shown in Fig. 4. 
The Fisher-Langer theory is a
great success for ferromagnets such as Fe\cite{fl_Fe} and Ni\cite{fl_Ni} 
and is reasonably successful for
antiferromagnets such as PrB$_6$\cite{fl_AF}. 
We already know from specific heat that bare electrons are renormalized 
to heavy electrons in CeRhIn$_5$\cite{hegger}, 
and from neutron diffraction that magnetic 
moment of Ce ion has been renormalized 
to 0.37$\mu_B$ at low temperatures\cite{bao00a}.
The Fisher-Langer behavior in Fig. 4 indicates that the Kondo interaction 
in CeRhIn$_5$ has been renormalized to the small value limit at low T. 

A fermion system weakly coupled to localized moments,
whose magnetic excitation spectra can be measured with inelastic neutron
scattering, is much easier for theoretical treatment than the Kondo 
lattice model. Established theory for superconductivity may be readily
applied to such a system. 
The Fisher-Langer behavior, therefore, may be used
as the indicator for heavy fermion materials which have
approached such a tractable fixed point.

\section{Summary}

Magnetic structures of CeRhIn$_5$ and Ce$_2$RhIn$_8$ have been determined
with neutron diffraction. Several phase transitions, including
commensurate-incommensurate transitions, are induced by magnetic
field applied in the basal plane. From neutron scattering 
measurement of the spatially dependent magnetic fluctuations and resistivities
measured in the basal plane and along the $c$ axis,
CeRhIn$_5$ is concluded to be a 3D system
both magnetically and in transport. The Fisher-Langer behavior in CeRhIn$_5$
suggests that the Kondo lattice of this material is renormalized to
heavy Fermi liquid weakly interacting with magnetic excitations
of the local moments. Critical behavior of magnetic order, which exists in some
heavy fermion materials, thus, can be used to probe the 
Kondo renormalization process.

\section*{Acknowledgments}
We thank C.M. Varma, S.M. Shapiro, C. Broholm, M.E. Zhitomirsky,
S.A. Trugman, J.M. Lawrence, A.V. Balatsky and D. Pines for useful discussions.
Work at LANL was support by US Department of Energy. 
ZF acknowledges NSF support at FSU, and PGP acknowledges FAPESP-Brazil.

\end{document}